\documentclass[aps,twocolumn,showpacs,superscriptaddress,floatfix,nofootinbib]{revtex4-2}

\usepackage{textcomp}
\usepackage{amsmath}
\usepackage{amssymb}
\usepackage{dsfont}
\usepackage{color}
\usepackage{graphicx}
\usepackage[normalem]{ulem}
\usepackage[linesnumbered,lined,vlined]{algorithm2e}
\usepackage[breaklinks=true,colorlinks,citecolor=blue,linkcolor=blue,urlcolor=blue]{hyperref}

\usepackage[final]{changes}

\begin{document}
\title{Many-body quantum sign structures as non-glassy Ising models}
\author{Tom Westerhout}
\email{tom.westerhout@ru.nl}
\affiliation{Institute for Molecules and Materials, Radboud University, Heyendaalseweg 135, 6525AJ Nijmegen, \mbox{The Netherlands}}

\author{Mikhail I. Katsnelson}
\email{m.katsnelson@science.ru.nl}
\affiliation{Institute for Molecules and Materials, Radboud University, Heyendaalseweg 135, 6525AJ Nijmegen, \mbox{The Netherlands}}

\author{Andrey A. Bagrov}
\email{a.bagrov@science.ru.nl}
\thanks{corresponding author}
\affiliation{Institute for Molecules and Materials, Radboud University, Heyendaalseweg 135, 6525AJ Nijmegen, \mbox{The Netherlands}}

\begin{abstract}
{\bf Abstract:} The non-trivial phase structure of the eigenstates of many-body quantum systems severely limits the applicability of quantum Monte Carlo, variational, and machine learning methods. Here, we study real-valued signful ground-state wave functions of frustrated quantum spin systems and, assuming that the tasks of finding wave function amplitudes and signs can be separated, show that the signs can be easily bootstrapped from the amplitudes. We map the problem of finding the sign structure to an auxiliary classical Ising model defined on a subset of the Hilbert space basis. We show that the Ising model does not exhibit significant frustrations even for highly frustrated parental quantum systems, and is solvable with a fully deterministic $O(K\log K)$-time combinatorial algorithm (where $K$ is the Ising model size). Given the ground state amplitudes, we reconstruct the signs of the ground states of several frustrated quantum models, thereby revealing the hidden simplicity of many-body sign structures.

\end{abstract}

\maketitle
\section*{Introduction}
The concept of phase of a quantum state is the main distinctive feature that makes the realm of quantum physics so different from the classical statistical mechanics~\cite{quantum_phase}.
Although it is unobservant itself, the wave function phase structure stands behind such phenomena as quantum interference~\cite{interference}, Anderson localization~\cite{localization}, many-body coherence~\cite{coherence}, and the Aharonov-Bohm nonlocality~\cite{Aharonov-Bohm,olariu}.
In many-body systems, the nodal surface, --- the surface of zeros of the wave function, --- encodes important information about the overall phase structure.
It is intimately connected to the entanglement properties of quantum states~\cite{Kaplis}.
For example, it has been demonstrated that real-valued non-negative wave functions cannot support extensive entanglement entropy scaling~\cite{Grover}.
Vice versa, long-range entangled quantum critical states are conjectured to have highly non-trivial (and even fractal) nodal surfaces~\cite{fractals} implying that the phase is a physically relevant property of quantum systems rather than a mere numerical abstraction.

On top of causing interesting and counter-intuitive emergent phenomena, non-trivial phase structures make studying interacting many-body systems notoriously difficult, especially when dealing with frustrations~\cite{frustrations} or finite-density fermionic matter~\cite{finite_density}.
Quantum Monte Carlo (QMC) algorithms suffer from the Sign Problem that spoils the sampling convergence~\cite{sign_problem} and represents the main barrier preventing researchers from routinely simulating generic quantum systems with classical hardware.
Resorting to variational wave functions, such as variational Monte Carlo ans\"atze~\cite{vmc}, tensor networks~\cite{tensor_networks}, or neural-network quantum states (NQS)~\cite{NQS_Carleo}, is one of the standard ways of getting around the Sign Problem.
For instance, convolutional neural networks have been successfully applied in combination with determinant QMC to identify phase transitions in many-body quantum systems suffering from sever forms of Sign Problem~\cite{Trebst}.
However, for the ground state search, optimizing the phases of many-body variational quantum states is a non-trivial problem either~\cite{sign_generalization, castelnovo} due to the excessive complexity in expressing phases~\cite{kastoryano} as well as the inability of the variational ansatz to generalize them from a limited sampled subset of the Hilbert space~\cite{sign_generalization}. 

On the other hand, if one is lucky enough to have a good approximation for the nodal surface or the phase structure of a quantum state, further analysis of the correlation effects becomes algorithmically straightforward~\cite{Ceperley}.
This principle underlies the fixed-node diffusion Monte Carlo~\cite{fixed_node} and path integral Monte Carlo algorithms~\cite{Ceperley_path}.
Using a proper nodal surface as a starting point of variational optimization schemes helps achieve good results for diverse physical systems such as spin liquids~\cite{spin_liquids, spin_liquids_2}, strongly correlated superconductors~\cite{Imada_SC}, and strange metals~\cite{strange_metals}.
This is one of the main reasons for success of some widely used variational ans\"atze such as the Slater-Jastrow~\cite{Slater_Jastrow_1, Slater_Jastrow_2} or the Gutzwiller-projected wave functions \cite{Gutzwiller, Nomura_fermions, neural_Gutzwiller}.

The utmost practical importance of retrieving phases of many-body quantum states
requires a deeper understanding of the Sign Problem and related structural properties of nodal surfaces. Computational complexity of the Sign Problem has been analyzed, and it was shown that, in its most general form, this problem is nondeterministic polynomially (NP) hard~\cite{NP_Troyer}. Its fundamental origins have been investigated and, within the framework of path integral QMC, have been related to topological properties of imaginary time evolution encoded in the analogue of the Aharonov-Anandan phase~\cite{Soluyanov}. Analytical insights into the geometry of the nodal surfaces of few-electron systems have also been obtained~\cite{nodal1, nodal2}.

Despite the generally high formal complexity of the Sign Problem, in each particular case, bypassing it could turn out to be an easy or a very difficult task.
Hence, it is natural to seek an approach for assessing the practical complexity of phase structures of generic many-body systems.
Here, we develop an alternative formal perspective on the problem of the reconstruction of many-body phase structures and relate its complexity to the complexity of combinatorial optimization of classical Ising models.
For that, we reverse the perspective of fixed-node approaches and ask a complementary question: if the knowledge of the phase (or nodal) structure helps the optimization algorithm so much, could the knowledge of the wave function amplitudes be used to efficiently reconstruct its phases?
Should this be true, amplitudes and phases could potentially be bootstrapped from each other in an algorithm such as variational Monte Carlo, where it has been shown that separating the tasks of optimizing amplitudes and phases leads to more accurate results~\cite{spin_liquids_2} (originally, the idea of separating amplitude and sign structures was introduced in \cite{sign_separating}).

In this paper, we address the problem of reconstructing many-body sign structures of ground states of time-reversal symmetric Hamiltonians. This class of systems embraces Heisenberg magnets with arbitrary couplings and types of anisotropy as well as Hubbard and extended Hubbard models on generic lattices, thus covering a plethora of phenomenologically important models. Specifically, we focus on several spin-$\frac12$ frustrated Heisenberg models. We show that, if amplitudes of a ground-state wave function are known, the sign structure can be reconstructed efficiently and with a high degree of accuracy. This can be done by mapping the task of sign structure optimization onto an auxiliary classical Ising model that is defined on the Hilbert space basis. We show that this auxiliary model is much less frustrated than the original quantum model and can be solved with a fast greedy algorithm. Although this paper focuses mainly on the specific problem of reconstructing the sign structures from amplitudes and leaves the task of implementing a complete variational ground state optimization algorithm for future studies, we demonstrate that the signs of the wave function coefficients can be obtained even if the amplitudes are known only up to a large error, and we provide a precise description of the complete variational algorithm in Supplementary Note 5.
\section*{Results}
\label{sec:Ising}
\begin{figure}[t!]
    \centering
    \includegraphics[width=0.95\columnwidth]{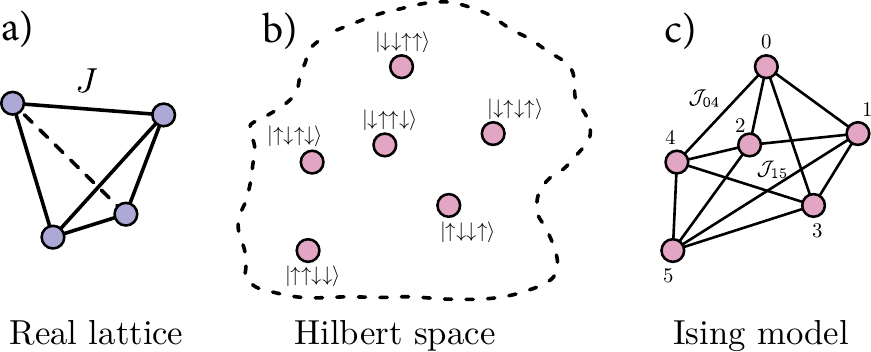}
    
    \caption{\label{fig:sign_ising}{\bf Cartoon illustration of the auxiliary classical Ising model for signs}. For the ground state of a quantum lattice model (a), the Hilbert space basis vectors with non-zero amplitudes (b) become sites of the classical Ising model (c). Couplings $\cal J$ between classical spins representing signs are defined by matrix elements of the quantum Hamiltonian and amplitudes of the basis vectors.}
\end{figure}
Consider a real-valued quantum lattice Hamiltonian $\hat{H}$. 
Its ground state can be expressed in the computational basis as
\begin{equation}
|\psi \rangle = \sum\limits_{i=1}^D \psi_i \, |i\rangle = \sum\limits_{i=1}^D \mathcal{S}_i |\psi_i| \,| i \rangle \,,
\label{eq:quantum_gs}
\end{equation}
where $\psi_i$ are real-valued, $\mathcal{S}_i = \mathrm{sign}(\psi_i)$, and $D$ denotes the Hilbert space dimension.
Energy of the ground state is given by the expectation value of $\hat{H}$:
\begin{equation}
    E = \langle \psi | \hat{H} | \psi \rangle = \sum\limits_{i,j = 1}^{D} \langle i | \hat{H} | j \rangle |\psi_i| |\psi_j| \, \mathcal{S}_i \mathcal{S}_j \,. \label{eq:gs_energy}
\end{equation}

Suppose, that the amplitudes $\{|\psi_i|\}_i$ are known, and the signs $\{\mathcal{S}_i\}_i$ are to be determined. Eq.~\eqref{eq:gs_energy} can be interpreted as the energy of a classical Ising model defined on the set of basis vectors of the Hilbert space, Fig. \ref{fig:sign_ising}:
\begin{equation}
    \mathcal{H} = \sum\limits_{i,j = 1}^D \mathcal{J}_{i,j} \mathcal{S}_i \mathcal{S}_j\,, \text{ where } \mathcal{J}_{i,j} = |\psi_i||\psi_j| \langle i | \hat{H} | j \rangle \,. \label{eq:Ising}
\end{equation}
The signs are binary variables, and the ground state sign structure is the ground state of the Ising model $\mathcal{H}$.
The Ising model is defined on the graph with topology determined by the quantum Hamiltonian $\hat{H}$ (when viewed as an adjacency matrix).

This simple mapping allows one to think about the complexity of quantum many-body sign structures in terms of the optimization complexity of classical spin models.
What are the conditions on $\mathcal{J}_{i,j}$ to have a reachable global minimum of Eq.~\eqref{eq:Ising}?
What is the difference between the auxiliary sign Ising models of simple and highly-frustrated quantum systems?
Do sign structures exhibit glassy behavior~\cite{glasses}?

As an illustration, let us start by considering an unfrustrated quantum Hamiltonian $\hat{H}$ such as the Heisenberg model on a square lattice with nearest-neighbor interactions only.
The absence of frustrations in the quantum model implies the absence of frustrations in the induced Ising model.
A proof of this statement for the Heisenberg model on an arbitrary lattice with or without anisotropy is provided in Supplementary Note 1, but we believe that the statement can be generalized to other types of interactions.
When the Ising model is unfrustrated, the sign structure can be obtained trivially with a greedy algorithm processing the graph in a depth-first or breadth-first manner~\cite{CLRS}.
Indeed, if some basis vector $|i\rangle$ has sign $\mathcal{S}_i$, the sign of its neighbor $|j\rangle$ is simply given by $\mathcal{S}_j = - \mathcal{S}_i \cdot \mbox{sign}(\langle i | \hat{H} |j\rangle)$.

Frustrated quantum systems are less trivial.
Nevertheless, we will show that, somewhat counter-intuitively, the auxiliary Ising model has a simple optimization landscape even when the parental quantum model is frustrated and has a complicated sign structure that is hard to obtain with traditional algorithms.
To this end, we first employ combinatorial optimization in a form of Simulated Annealing (SA) algorithm~\cite{SA}, and then show that good results can be obtained even with an $\mathcal{O}(K \log K)$ greedy algorithm, where $K$ is the size of the Ising model.
Combinatorial optimization has been employed earlier in an attempt to learn the sign structures of some frustrated magnets~\cite{Nikita_sign}. However, in that paper, genetic optimization was used for the training of a neural network whereas here we directly optimize the sign structure.

In the following, we will demonstrate that

\begin{enumerate}
    \item For relatively small quantum systems, SA finds exact ground states of the auxiliary Ising models with high probability, and the greedy algorithm even manages to find them in a deterministic way.
    The Hilbert space dimension is still large enough to claim that classical glasses of such size are not amenable to global optimization.
    \item For larger quantum systems where global optimization is unfeasible, local optimization on small connected clusters of the Ising models yields a very good approximation to the global solution.
\end{enumerate}
Together, these two findings indicate that the auxiliary Ising model is frustrated very mildly, and that the complexity of the sign structure identification is far below that of finding the ground state of a classical glass, at least in a number of non-trivial physically relevant cases.

\subsection*{Small quantum systems}
\begin{figure}[t!]
    \centering
    \includegraphics[width=\columnwidth]{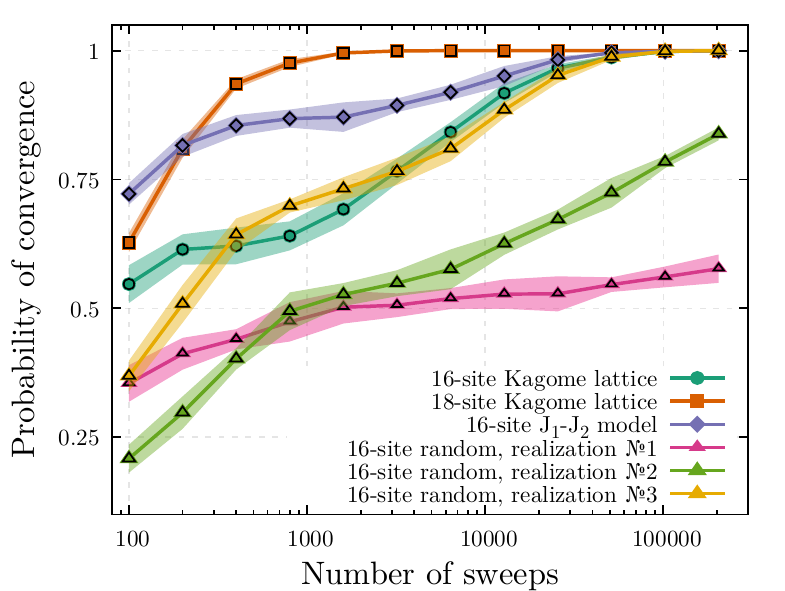}
    
    \caption{\label{fig:small_models}{\bf Quality of simulated annealing optimization of the sign structure for small quantum models.}
    We show the probability of convergence to at least $99.5\%$ accuracy as a function of the number of sweeps.
    Solid points are the numerical data, and shaded regions indicate the uncertainty (plus-minus two standard deviations).
    Probability of convergence is obtained by running the simulated annealing algorithm 1024 times and counting the number of converged runs.
    To estimate the uncertainties, we repeat the simulation 10 times.
    So in total, 10240 simulated annealing runs were performed to obtain each point.
    For frustrated models with regular lattice geometries, the probability of convergence approaches $1$ when the number of sweeps is increased.
    For the random fully-connected model, it depends on the concrete realization of couplings, but still exceed $50\%$.}
\end{figure}

Let us start by considering a few small clusters: antiferromagnetic Heisenberg model on 16 and 18-site Kagome lattices, $J_1$-$J_2$ model on 16-site square lattice with the nearest-neighbour ($J_1$) and next-nearest-neighbour ($J_2$) interactions, and 16-site Heisenberg model with random all-to-all exchange interaction.
For the $J_1$-$J_2$ model, we focus on the case when $J_2 = 0.55 J_1$ which is considered to be one of the most difficult points for variational algorithms.
For the random Heisenberg model we draw exchange couplings $J_{i,j}$ from the Gaussian distribution with zero mean and variance 1. More details about the systems are provided in Supplementary Note 2.

These models are easily solvable with the exact diagonalization (ED) method, and their ground states belong to the sectors of zero magnetization.
The resulting dimensions of the auxiliary Ising models are then
\begin{equation*}
    D = \binom{N}{N/2} 
      = \left\{
      \begin{aligned}
         &12870\mbox{, when }N=16, \\
         &48620\mbox{, when }N=18
      \end{aligned}
      \right.
\end{equation*}
If these sign Ising models had glassy energy landscapes, the task of finding their ground states would be far beyond the capacity of SA or devices such as D-Wave quantum annealers (Advantage\texttrademark\ QPU, for instance, can handle more than 5000 variables, but is limited to a special connectivity-15 graph~\cite{boothby2020next}).

In Fig.~\ref{fig:small_models} we show the probability of convergence for the Simulated Annealing algorithm (parameters of the algorithms are provided in Supplementary Note 3).
We say that the algorithm has converged if it has reached at least $99.5\%$ accuracy.
This threshold is chosen, on one hand, to avoid problems arising from numerical inaccuracies --- some basis vectors have zero amplitudes up to machine precision, and their signs cannot be determined faithfully.
On the other hand, the threshold is still high enough to consider the results exact for all practical purposes.

\begin{figure}[t]
    \centering
    \includegraphics[width=0.9\columnwidth]{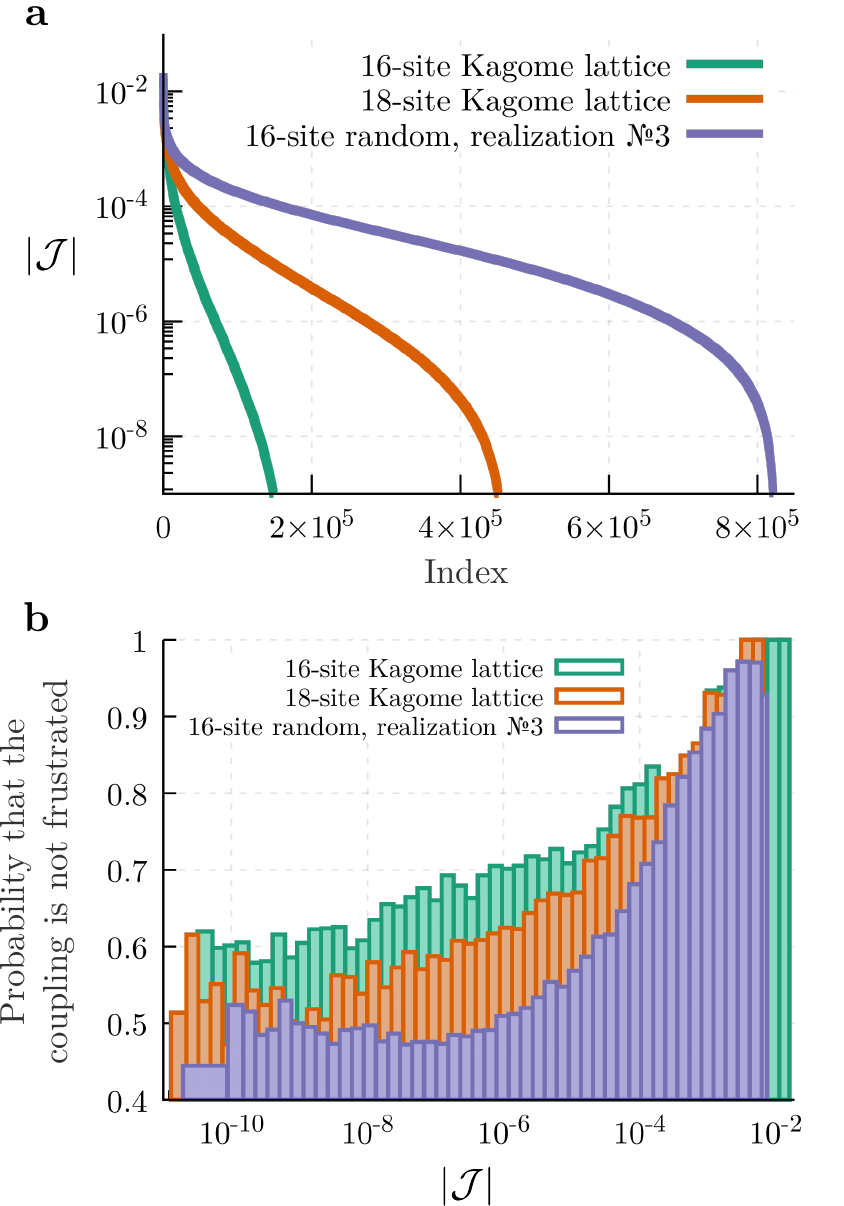}
    \caption{\label{fig:coupling_distribution}{\bf Distribution of the Ising model couplings.} \textbf{a} Sorted distribution of couplings of the Ising models corresponding to ground states of the studied quantum systems. In the logarithmic scale it is evident that large couplings comprise only a small fraction within the whole graph and, hence, could be sparsely distributed. \textbf{b} Histogram of the probability that a coupling of given magnitude is not frustrated (in other words, the locally optimal state of the corresponding two spins is compatible with the global solution). Larger couplings are likely not frustrated, which underlies the success of simple optimization algorithms.}
\end{figure}

It turns out that for the Heisenberg model on the Kagome lattices and for the $J_1$-$J_2$ model on the square lattice, the algorithm almost always converges if the number of Simulated Annealing sweeps is at least $50000$.
For the random all-to-all connected models, the chance of finding the exact sign structure depends on the particular realization, but still exceeds $50\%$ when the number of sweeps is at least $10000$.
Hence, repeating SA 20 times (which can be done in parallel) with $10000$ sweeps ensures that we find the ground state with more than $99.9999\%$ probability.
These results should be compared to Fig.~4 of Ref.~\cite{heim2015quantum} where SA is applied to spin glasses with $6400$ spins, and even after $10^8$ Monte Carlo steps, the residual energy is still above $10^{-3}$.
Our experiments thus show that, for rather generic and even random bi-local frustrated quantum systems, the sign optimization landscape is very mildly frustrated.

To understand the origin of such non-glassy behavior, it is instructive to take a look at the distributions of couplings of the auxiliary Ising models. They arises from the distributions of amplitudes $|\psi_i|$ of the corresponding quantum Hamiltonian ground states, which are often strongly peaked with a relatively small fraction of large amplitudes.
In turn, the distributions of $|{\cal J}_{i,j}|$ are also peaked as shown in Fig.~\ref{fig:coupling_distribution}a. As a result, the model landscapes are likely dominated by a moderate number of strong couplings that are sparsely distributed across the Ising model graphs and do not strongly compete with each other.
If it is truly the case, there is a possibility that a nearly optimal solution of the Ising model can be obtained even without full combinatorial optimization.
As can be seen from Fig.~\ref{fig:coupling_distribution}b, the optimal sign structure does not induce frustration on most links with large values of $|{\cal J}_{i,j}|$.
To further explore this, we attempt to reproduce the exact solution by using a simple greedy algorithm that goes through each of the Ising spins only once and performs local optimization.
The algorithm processes couplings in the decreasing order of $|\mathcal{J}_{i,j}|$ and relies on simple heuristics to minimize frustrations.
A detailed description of the algorithm is provided in the Methods section.
Results of executing this algorithm on the 16- and 18-spin models are given in Tab.~\ref{tab:greedy}.
One can see that for the $J_1$-$J_2$ model, one of the Kagome lattices, and two instances of the all-to-all connected model with random couplings, our completely deterministic local optimization procedure gives exact results, and for the remaining models --- imperfect, but high quality solutions.
In contrast with SA, this algorithm is immanently non-stochastic and always produces the same solution, so there is no need to do many sweeps or run the algorithm a number of times to collect statistics.
This indicates that frustrations in the auxiliary Ising models truly are almost negligible.

\begin{table}[t!]
\centering
\begin{tabular}{|l|c|c|} 
 \hline
 System & Accuracy & Overlap \\
 % [0.9ex] 
 \hline\hline
 16-site $J_1$-$J_2$ model & 1.0 & 1.0  \\ 
 16-site Kagome lattice & 1.0 & 1.0  \\
 18-site Kagome lattice & 0.998 & 1.0  \\
 16-site random, \textnumero 1 & 1.0 & 1.0  \\
 16-site random, \textnumero 2 & 1.0 & 1.0  \\
 16-site random, \textnumero 3 & 0.945 & 0.885  \\
 % random \textnumero 3 & 1.0 & 1.0  \\ [1ex] 
 \hline
\end{tabular}
\caption{{\bf Results of the greedy optimization for small quantum systems.} The simulations are fully deterministic. Accuracy and overlap are computed on the full Hilbert space.}
\label{tab:greedy}
\end{table}

\begin{figure}[t]
    \centering
    \includegraphics[width=\columnwidth]{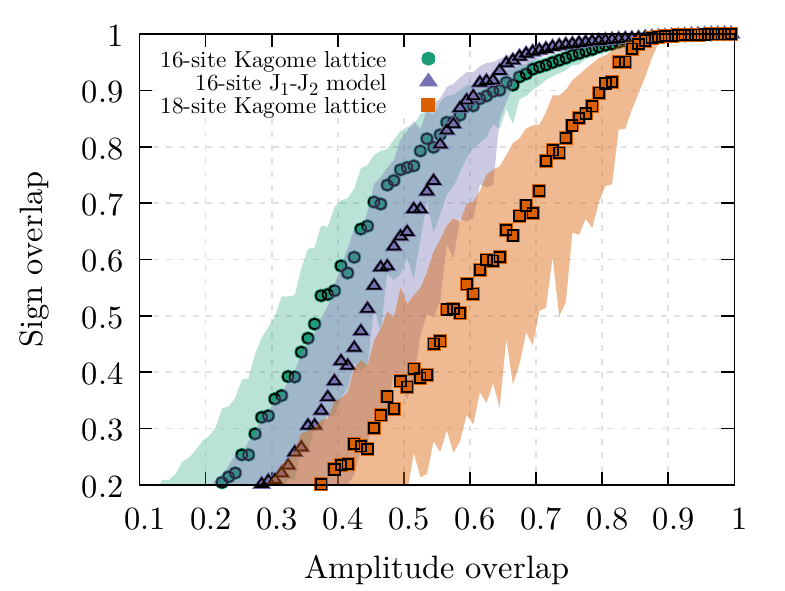}
    \caption{\label{fig:noise_small}{\bf Robustness of the retrieved sign structure.} Quality of the sign structure measured by the sign overlap $O^{\cal S}$ as a function of the quality of amplitudes measured by the amplitude overlap $O^{\cal A}$. For regular lattices, $O^{\cal S}$ stays above $0.9$ as long as $O^{\cal A}\gtrsim 0.8$. Solid points are median values after running the optimization for a 100 times with a given $\epsilon$. Shades depict interquartile ranges.}
\end{figure}

Importantly, this sign structure is not only retrievable with high accuracy in $\mathcal{O}(K \log K)$ time (where $K$ is the number of Ising spins), but can be shown to be robust to noise in the amplitudes of the wave function. To this end, we corrupt amplitudes of the ground state by adding noise to the logarithms of the amplitudes:
\begin{equation}
    \log \left|\psi^\mathrm{corrupt}_i\right| \equiv \log \left|\psi^\mathrm{exact}_i\right| + \epsilon_i, \,\,\, \epsilon_i \sim \mbox{Uniform}(-\epsilon, \epsilon ), 
\end{equation}
and vary the maximal noise amplitude $\epsilon$. To gauge the overall magnitude of noise, we introduce a measure that we call amplitude overlap between the exact and the corrupted states:
\begin{equation}
    O^{\cal A} \equiv \sum_i |\psi^\mathrm{exact}_i|\cdot |\psi^\mathrm{corrupt}_i|,
\end{equation}
where both the exact and corrupted wave functions are normalized.
Without the noise ${\cal O}^{\cal A} = 1$, and smaller values of the amplitude overlap correspond to higher levels of noise.
In particular, if $\psi^\mathrm{corrupt}_i$ are fully random and bear no information about the true solution, the amplitude overlap approaches $0.3$--$0.5$ for the models from Tab.~\ref{tab:greedy}. Results for small regular lattices are shown in Fig. \ref{fig:noise_small}, where quality of the retrieved sign structure is measured by the sign overlap that is defined as
\begin{equation}
    O^{\cal S} \equiv  \sum_i |\psi^\mathrm{exact}_i|^2 {\cal S}_i^\mathrm{exact}{\cal S}_i^\mathrm{retrieved}
\end{equation}
One can see that the sign structure remains very close to the exact one for $O^{\cal A} > 0.8$.
For example, for the 16-site Kagome lattice, $O^\mathcal{A} \approx 0.8$ corresponds to $\epsilon \approx 1.45$.
The maximal amplitude of the exact ground state is $\max_i |\psi^\mathrm{exact}_i| = 0.0924$, and its corrupted values fall into the range $[0.02, 0.4]$.
At first glance, it seems unbelievable that such extreme deformations of amplitudes have mild effect on the sign structure.
A possible explanation of this phenomenon is that, due to the peaked nature of the distribution of amplitudes, even strong noise in the amplitudes does not significantly change their order: if $|\psi^\mathrm{exact}_i| \gg |\psi^\mathrm{exact}_j|$, their corrupted counterparts would likely still satisfy $|\psi^\mathrm{corrupt}_i| > |\psi^\mathrm{corrupt}_j|$.
The reconstructed sign structure changes very little if this order is not severely altered, irrespective of the concrete values of amplitudes.
For random all-to-all connected lattices, the robustness of the sign structure is less striking, but still pronounced.

\subsection*{Finding sign structures of large quantum systems}
In the real cases of interest, the Hilbert space dimension would be too large to perform optimization of signs globally on the full basis. Hence, we need to make sure that it is possible to find the sign structure via local optimization on a subset of the basis. Given the efficiency of the greedy algorithm in the case of small systems, we use it for all numerical experiments in this section and provide some results of executing SA in Supplementary Note 6.

Let $\cal C$ be a connected subset of the basis. Ising spins within $\cal C$ are connected to each other and to ``external'' spins associated with Hilbert space vectors outside of $\cal C$. The Ising Hamiltonian \eqref{eq:Ising} can then be split into three parts
\begin{equation}
    \mathcal{H} = \sum\limits_{i,j \in {\cal C}} {\cal J}_{ij} \mathcal{S}_i \mathcal{S}_j + \sum\limits_{n \in \partial {\cal C}} \mathcal{S}_n\sum\limits_{k \notin {\cal C}} {\cal J}_{nk} \mathcal{S}_k + \text{const},
\end{equation}
where $\partial C$ denotes the spins in $\cal C$ that have at least one external connection.
The energy contribution from the connections not involving $\mathcal{C}$ is treated as a constant.
If the greedy algorithm is used to perform the optimization on $\cal C$, it does not change the state of spins outside of the subset, $\mathcal{S}_k, \,\, k \notin \cal C$, and they can be regarded as external magnetic fields $h_n=\sum\limits_{k \notin {\cal C}} {\cal J}_{nk} S_k$.
If $h_n$ are known, the optimization problem can be solved exactly when $\cal C$ has a reasonable size (up to a few million elements).
However, in reality, the states of the spins outside of $\mathcal{C}$ are unknown, and one needs to make some assumptions about them.
The simplest way to proceed is to impose $h_n=0$ and optimize the signs in $\cal C$ by taking into account just the internal connections. This approximation is valid only if the cumulative strength of the internal connections is much higher than that of the external fields.

\begin{figure}[t]
    \centering
    \includegraphics[width=0.7\columnwidth]{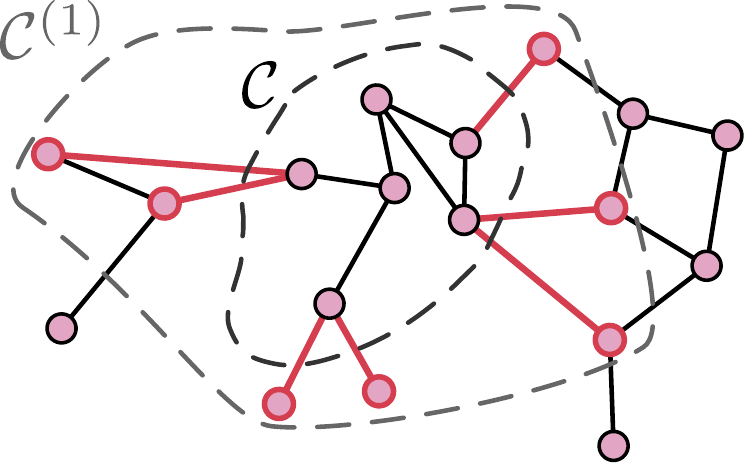}
    \caption{\label{fig:extension}{\bf The Hamiltonian extension idea.} We start with a connected cluster $\cal C$ of Ising spins. To obtain its extension ${\cal C}^{(1)}$, we additionally include all spins that have at least one connection with $\mathcal{C}$. These spins and the corresponding connections are highlighted in red.}
\end{figure}

\begin{figure*}[t]
    \centering
    \includegraphics[width=0.95\textwidth]{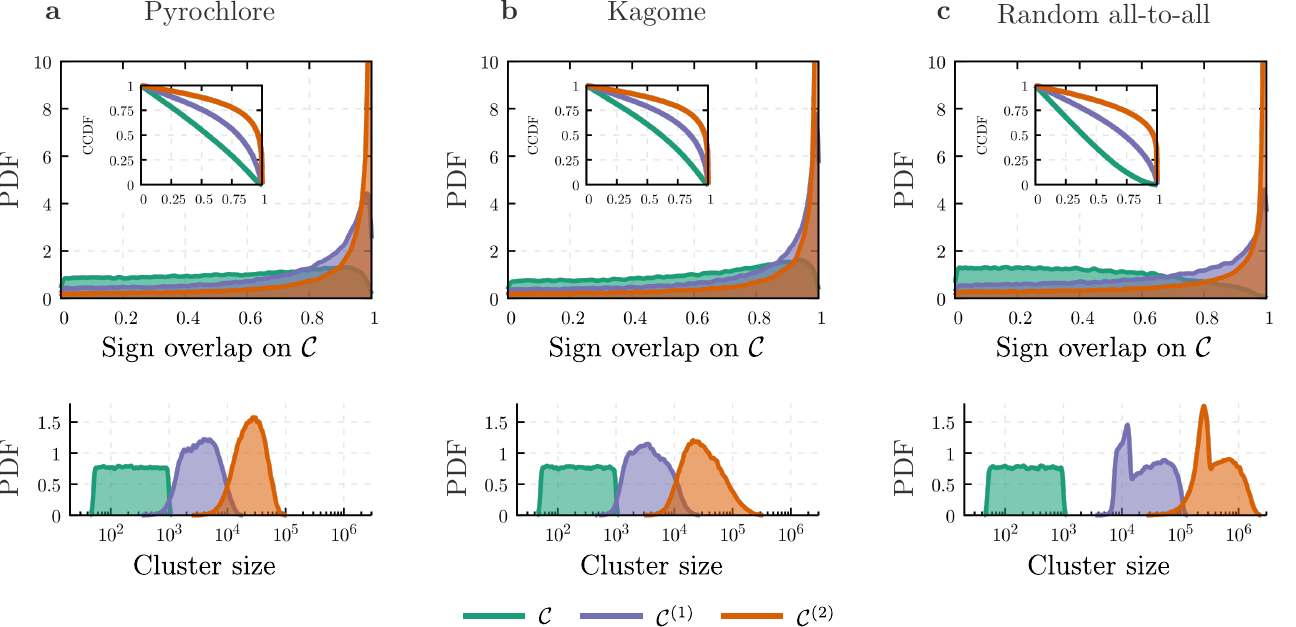}
    \caption{\label{fig:overlap_and_clusters}{\bf Quality of the sign structures on random small connected clusters from the greedy algorithm.} The results are shown for \textbf{a} the 32-site pyrochlore lattice, \textbf{b} 36-site Kagome lattice, and \textbf{c} 32-site all-to-all connected model with random couplings. 
    The top panel shows the distribution (or the probability density function, PDF) of clusters as a function of the resulting normalized sign overlap $O^{\cal S}_\mathcal{C}$. We run the greedy algorithm on the original cluster $\mathcal{C}$ as well as on its Hamiltonian extensions ${\cal C}^{(1)}$ and ${\cal C}^{(2)}$. The insets show complementary cumulative distribution functions (CCDF) showing the overall probability to obtain normalized sign overlap higher than a chosen value. The bottom panel shows how sizes of the clusters and their extensions are distributed. Subfigure \textbf{a} is plotted using $100\,000$ data points, subfigures \textbf{b} and \textbf{c} -- using $80\,000$ data points.}
\end{figure*}

This property can be ensured with a simple trick. Spins $\mathcal{S}_i$ and $\mathcal{S}_j$ of the classical Ising model are connected if the corresponding quantum Hamiltonian matrix element is nonzero: $\langle i | H | j \rangle = \mathcal{J}_{ij} \neq 0$. One can construct a Hamiltonian extension ${\cal C}^{(1)}$ of the cluster $\cal C$ by adding all the external sites $\mathcal{S}_k \notin {\cal C}$ such that $\mathcal{J}_{kj} \neq 0$, for $\mathcal{S}_j \in \partial {\cal C}$. 
That way, each spin $\mathcal{S}_j$ that previously resided on the boundary $\partial \mathcal{C}$ now belongs to the bulk.
Fig.~\ref{fig:extension} illustrates this procedure.
Higher order extensions ${\cal C}^{(i)}$ can be obtained similarly.
After two or three steps, one obtains a large cluster that is dominated by internal connections, and $h_n$ can be safely neglected when running combinatorial/greedy optimization on the extended cluster.
The original spins $\cal C$ lie deep inside the larger cluster, and their signs can be reconstructed with good accuracy. It should be noted that, due to the sparsity of the quantum Hamiltonian, these extensions would be of much smaller size than the dimension of the Hilbert space even in the case of all-to-all connected Heisenberg model (see the bottom panel of Fig.~\ref{fig:overlap_and_clusters}).

To test the above procedure, we consider the ground states of the Heisenberg model on the 36-site Kagome lattice \cite{kagome36} and 32-site pyrochlore lattice (the $2\times 2\times 2$ configuration) \cite{spin_liquids_2}, and one realization of the random Heisenberg model with all-to-all connections on 32 sites.
All these models can still be solved with ED, but the corresponding Hilbert spaces become large enough to serve as a test bed for studying the quality of the greedy optimization on smaller subsets.
For the random Heisenberg model, the only symmetry we use is $U(1)$ (magnetization conservation), and the Hilbert space dimension of the zero-magnetization sector is $D = \binom{32}{16} \approx 6\cdot 10^8$. For the pyrochlore and Kagome lattices, lattice symmetries can be taken into account, and the corresponding Hilbert space dimensions are approximately $8 \cdot 10^5$ and $3 \cdot 10^7$, correspondingly.

For each ground state, we sample random connected clusters $\cal C$ of various sizes.
For each of them, we construct extensions of the first and second orders, and run greedy optimization on both the original $\cal C$ and the extended ${\cal C}^{(i)}$ clusters ignoring the external fields.
The quality of the obtained sign structure is measured by the normalized sign overlap between the variational wave function (with signs from the greedy optimization and amplitudes from ED) and the true one (with both signs and amplitudes from ED), which we compute on the basis states belonging to $\cal C$:
\begin{equation}
    O^{\cal S}_{\cal C} = \frac{1}{\sum_{i \in {\cal C}} |\psi_i|^2}
        \cdot \sum\limits_{i \in {\cal C}}
            \overbrace{|\psi_i|\mathcal{S}_i}^\text{variational}
            \cdot \underbrace{\psi_i}_\text{exact} \,,
\end{equation}
where $\mathcal{S}_i$ are obtained from the greedy optimization.

In Fig.~\ref{fig:overlap_and_clusters}, we show the quality of the greedy optimization for \textbf{a} the 32-site pyrochlore lattice, \textbf{b} the 36-site Kagome lattice, and \textbf{c} the 32-site random Heisenberg model.
The overlap $O^{\cal S}_\mathcal{C}$ depends on the extension order, and, as can be seen from the complementary cumulative distribution functions, constructing the second order extension appears sufficient to reach $O^{\cal S}_\mathcal{C} \simeq 0.9$ with a chance of about 70\% on all systems. Clusters $\cal C$ were sampled in a way to ensure that their sizes are logarithmically distributed in the range $[50,1000]$, which is a typical range of sizes that one would encounter in the context of variational Monte Carlo as described in Supplementary Note 5.
There, we also analyze how the quality of the obtained sign structures is affected by noise in amplitudes.
Just as for the smaller systems, the algorithm is robust to noise (Supplemental Note 4), but for optimal results, the optimization should be carried out on the third-order extensions ${\mathcal C}^{(3)}$.

\begin{figure}[t]
    \centering
    \includegraphics[width=0.9\columnwidth]{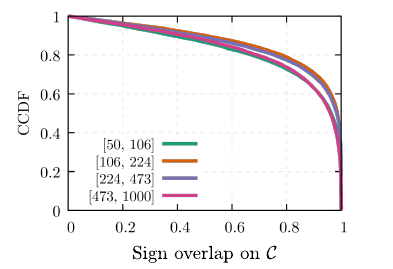}
    \caption{\label{fig:pdf_sizes}{\bf Sign structure quality and the cluster $\cal C$ size do not correlate.} Complementary cumulative distribution functions (CCDF) of sign overlaps for different ranges of the original cluster $\cal C$ sizes (as shown in the legend) for the 36-spin Kagome model. Similarity of the curves clearly indicates that correlation between quality of optimization and cluster sizes is insignificant.}
\end{figure}

Fig.~\ref{fig:pdf_sizes} shows that there is no real dependence of the quality of optimization on the original cluster size: probabilities of having the normalized sign overlap $O^{\cal S}_{\cal C} \gtrsim 0.9$ are approximately the same for the clusters of sizes $[50, 106]$, $[106, 224]$, $[224, 473]$, and $[473, 1000]$.

This, together with the fact that ${\cal C}^{(2)}$ extensions are much smaller than the Hilbert space basis, Fig. \ref{fig:overlap_and_clusters} (bottom panel), tells us that the result of local cluster optimization is compatible with the global sign structure. In other words, the Ising model is not glassy. 

\section*{Discussion}
In this paper, we have suggested a way of analyzing the sign structure of a many-body quantum system ground state through the lens of an auxiliary classical Ising model.
The Ising model is defined by the quantum Hamiltonian and the amplitudes of the ground state. We have shown that such models turn out to be mildly frustrated and can be easily solved with the help of standard combinatorial optimization algorithms such as Simulated Annealing or even a simple greedy algorithm.

A naturally arising question is how having a close-to-optimal sign structure on a small subset of the Hilbert space basis helps in solving any practical problems. The most straightforward way to make use of the suggested approach is to incorporate it into the the context of variational Monte Carlo. There, one never accesses the complete Hilbert space. At every step of optimization, energy of the trial state and its gradient with respect to the variational parameters are computed based on Monte Carlo sampling. The sign structure thus
only needs to be optimized on small set of nodes of the auxiliary Ising model that scales as $\mathcal{O}(N)$ with the size of the quantum system (see Supplementary Note 5 for the detailed explanation).
Using the variational representation to only encode amplitudes and having explicit representation for the signs of the relevant basis vectors, one can hybridize the variational optimization of the amplitude structure with greedy or combinatorial sign optimization.
Since the greedy algorithm is log-linear in time, complexity of the sign structure optimization step within the variational loop then scales as $\mathcal{O}(N^4\log N)$ with the number of quantum spins if one is using the second-order extensions, as shown in Supplementary Note 5.
Although the implementation of the complete algorithm goes beyond the scope of this paper, in Supplementary Note 5 we describe it in detail, and explain why nuances such as the necessity of optimization of disconnected clusters, possibly large number of relevant vectors, as well as noise unavoidable in stochastic approaches, do not diminish the applicability of this approach.

In particular, if successfully implemented, the combinatorial approach to sign reconstruction could aid the optimization of NQS representations. 
Since the signs can be bootstrapped directly from the amplitudes, they would not need to be encoded in a neural network at all, and the typical issues with learning the sign structure, such as the sign structure generalization problem~\cite{sign_generalization}, simply would not arise.
For example, it is instructive to compare the quality of sign structures obtained from mapping on the auxiliary Ising model with what has been achieved by training simple dense neural networks in a conceptually similar setting~\cite{sign_generalization} (exactly known amplitudes, optimization of signs).
In the current approach, for the 36-site Kagome lattice, we routinely get partial sign overlap of above $90\%$. In the prior studies~\cite{sign_generalization}, for the 30-site Kagome lattice, the neural networks could not even reach a $40\%$ sign overlap.
Moreover, here we sample clusters from a nearly uniform probability distribution $p(i) \sim |\psi^i|^{1/10}$ to ensure that signs of basis vectors with both large and small amplitudes are reconstructed correctly (see the Methods section).
The NQS, on the other hand, fails to properly learn even the signs of the ``dominant'' basis vectors with large amplitudes sampled from $p(i) \sim |\psi_i|^2$.

There are a few directions in which our work can be extended.
We outline the ideas, but leave the implementation for future studies because we would like to keep the focus of the current paper on the introduced concepts.
First of all, the way we treat the sign structures on the extended clusters is suboptimal. For instance, we make the simplest possible approximation that the magnetic field values at the boundary of the extensions are $h_n=0$. However, one can think of designing a non-trivial self-consistent approximation for them, somewhat in the spirit of Dynamical Mean Field Theory~\cite{DMFT} that would systematically improve the local sign structure by iterative adjustment of $h_n$.
Besides that, although the classical greedy algorithm is extremely fast, the combinatorial optimization might lead to more accurate results. Since it is much more time consuming, one can think of making it more efficient by executing it on hardware quantum annealers~\cite{QA,QA2,QA3}, and although the extended clusters might be too big for such devices at the moment, there are ways to overcome this limitation~\cite{raymond2022hybrid,okada2019improving}.

In this paper, we have focused on quantum systems with binary sign structures.
In general, when the Hamiltonian is complex-valued, the phases can be mapped onto an XY model~\cite{XY_model} as shown in Supplementary Note 7.
Finally, although we have studied only the ground states, the techniques are applicable to other problems for which a variational principle can be formulated. Searching for excited states~\cite{excited} and simulating the time evolution~\cite{TDVP} of a wave function are two examples.

\section*{Methods}
In this paper, we used exact ground state wave functions for quantum spin systems of up to 36 sites as the ground truth reference states, which we obtained with the SpinED package~\cite{SpinED}.

For systems of 16 or 18 spins, we determined the coefficients of the Ising model~\eqref{eq:Ising} and ran both SA and the greedy algorithm on the full Hilbert space graph to optimize the sign structure.
The parameters of the SA algorithm are provided in Supplementary Note 3.
The greedy algorithm is outlined in this section.

For larger systems of 32--36 spins, we first sampled small connected clusters $\cal C$ of around 100--1000 basis vectors, constructed their Hamiltonian extensions ${\cal C}^{(1)}$ and ${\cal C}^{(2)}$, and ran the greedy algorithm to optimize the signs on the extensions.
Some results of running the SA algorithm are provided for comparison in Supplementary Note 6.

\subsection*{Constructing small connected clusters within large Ising graphs}

To analyze whether the ground state sign structures were amenable to local optimization, we performed simulations on systems that have a large Hilbert space, but could still be diagonalized exactly.
For each system, we have studied how well the greedy algorithm was able to predict the signs on random connected clusters.

The clusters were constructed by choosing a random starting point and then extending the cluster until it reached the desired size.
The initial points were obtained by direct sampling from $|\psi|^{1/10}$.
This ensured that basis states with both high and low amplitudes were covered, but zero amplitudes (up to numerical noise) were not considered since even ED might not be accurate around those points.
Algorithm~\ref{alg:random-cluster} shows how the initial cluster was extended until it reached the required size.

\normalem
\begin{algorithm}[H]\label{alg:random-cluster}
    \DontPrintSemicolon
    \KwData{$x$ --- initial point around which the cluster is built;
            $N$ --- size of the cluster; $p$ --- probability of keeping a site;
            $\mathcal{J}_{ij}$ --- couplings of the auxiliary Ising model.}
    \KwResult{Connected cluster $\mathcal{C}$ of size $N$.}
        $\mathcal{C} \longleftarrow \{ x \}$\;
    \While{$\# \mathcal{C} < N$}{
        $js \longleftarrow \{ j \;|\; \mathcal{J}_{ij} \neq 0, \text{ where } i \in \mathcal{C},\; j \in \mathcal{H} \}$ \;
        \For{$j \in js$}{
            $u \longleftarrow \mathrm{Uniform}(0, 1)$\;
            \If{$u < p$}{
                $\mathcal{C} \longleftarrow \mathcal{C} \cup \{ j \}$\;
                \If{$\# \mathcal{C} = N$}{\textbf{break}\;}
            }
        }
    }
\caption{Create a random cluster around the specified point $x$. In all our simulations, the parameter $p$ was set to $1/2$.}
\end{algorithm}
\ULforem

With the cluster $\mathcal{C}$ at hand, we ran the greedy algorithm on it as well as on its Hamiltonian extensions $\mathcal{C}^{(1)}$ and $\mathcal{C}^{(2)}$ to predict the wave function signs on $\mathcal{C}$.
The overlap with the exact ground state, computed on the basis vectors belonging to the cluster, was used to measure the quality of our optimization procedure.

\begin{figure}[h]
    \centering
    \includegraphics[width=0.65\columnwidth]{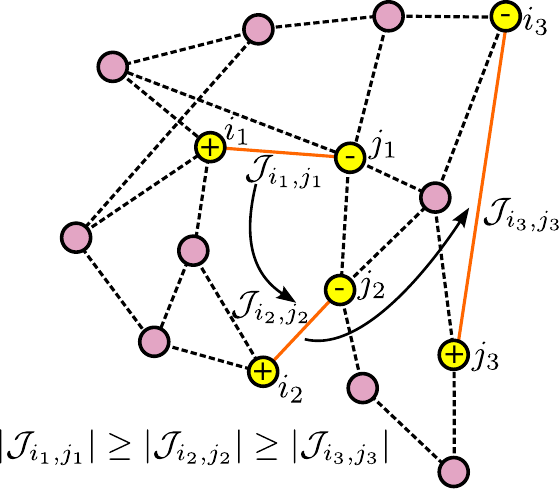}
    \caption{{\bf The idea of the greedy algorithm.} The greedy algorithm first locally optimizes signs connected by strongest couplings.}
    \label{fig:greedy}
\end{figure}

\begin{figure*}[t]
    \centering
    \includegraphics[width=0.8\textwidth]{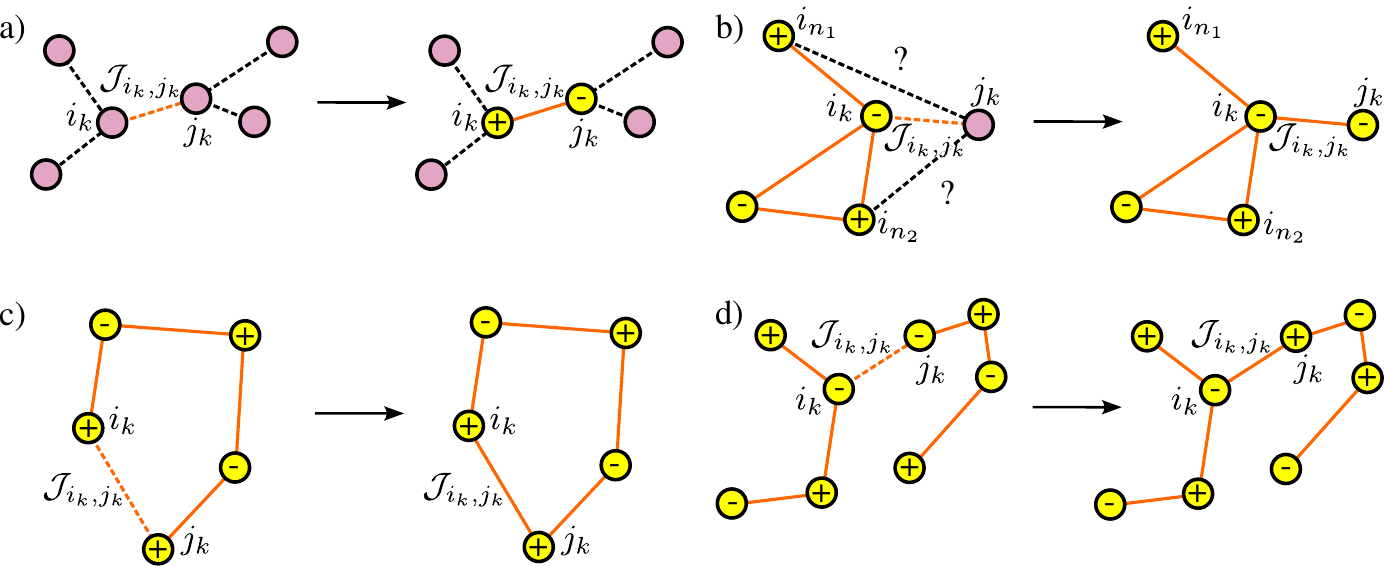}%width=1\columnwidth]
    \caption{\textbf{Possible decisions the greedy algorithm makes at every step depending on the states of sites and configurations of already scanned clusters.} Yellow nodes and orange edges belong to $G'$ graph that is being constructed during the algorithm evaluation time.}
    \label{fig:greedy_steps}
\end{figure*}

\subsection*{Deterministic greedy algorithm for signs reconstruction}
In this section we describe the greedy algorithm for sign structure optimization on a connected sub-graph $G$ of the Ising model, where $G$ can span the entire Hilbert space or a ${\cal C}^{(n)}$ extension of some cluster of interest.
Each edge $k$ of the graph is a tuple $(i_k, j_k, \mathcal{J}_{i_k, j_k})$ where $i_k$ and $j_k$ are indices of Ising spins connected by the edge $k$.

The idea is to try to minimize the local energies of the strongest edges, Fig. \ref{fig:greedy}.
We iterate over edges in decreasing order of their $|\mathcal{J}_{i_k,j_k}|$ and add them to an initially empty graph $G'$.
In other words, all nodes in $G'$ have their signs assigned and all nodes in $G \setminus G'$ --- not yet.
For each edge, we use the following heuristic for choosing the signs:

\begin{itemize}
    \item If both $i_k$ and $j_k$ are in $G\setminus G'$ (i.e.,  do not yet have signs), we choose $\mathcal{S}_{i_k}$ and $\mathcal{S}_{j_k}$ such that the energy is minimized, or, in other words, such that ${\cal J}_{i_k,j_k} \mathcal{S}_{i_k} \mathcal{S}_{j_k} < 0$ (See Fig.~\ref{fig:greedy_steps}a).
    
    \item If only one of the signs, say, ${\cal S}_{i_k}$ is already assigned, we choose ${\cal S}_{j_k}$ to minimize the energy, i.e. such that
    \begin{equation*}
        \mathcal{S}_{j_k}
            \sum_{i_n} {\cal S}_{i_n} \mathcal{J}_{i_n,j_k} < 0
    \end{equation*}
    where the sum runs only over $i_n$ that belong to the same connected component of $G'$.
    Note that edges $(i_n, j_k, \mathcal{J}_{i_n, j_k})$ do not yet belong to $G'$ (See Fig.~\ref{fig:greedy_steps}b).

    \item If both signs are assigned and belong to the same connected component of $G'$, we do nothing (See Fig.~\ref{fig:greedy_steps}c).    
    
    Motivation: Since we process edges in the decreasing order of their coupling strengths, the coupling $\mathcal{J}_{i_k, j_k}$ cannot be stronger than any of the couplings in the connected component. Hence, even the edge turns out to be frustrated, it should not be optimized as flipping the sign of either $i_k$ or $j_k$ would increase the energy of the part of the Ising model that we have already processed.
    
    \item If both signs are assigned and belong to different connected components of $G'$, we merge the two components flipping, if necessary, all signs in the second one (when $\mathcal{S}_{i_k} \mathcal{S}_{j_k} \mathcal{J}_{i_k,j_k} > 0$).
    This again minimizes the energy of the already processed part of the Ising model (See Fig.~\ref{fig:greedy_steps}d).
\end{itemize}

\textbf{Data availability:} Numerical data are stored in repository~\cite{zenodo-data}, and the scripts used to generate them are stored in repositories ~\cite{SA-code-1, zenodo-code}.
\vskip 10pt
\textbf{Code availability:} Exact diagonalization has been performed using the SpinED package~\cite{SpinED}.
We have used the lattice-symmetries package~\cite{lattice-symmetries} for all other manipulations with many-body wave functions.
The Simulated Annealing and greedy algorithms are implemented in the ising-glass-annealer package~\cite{SA-code-2}.
All other scripts that have been used to produce and process the data can be found in~\cite{SA-code-1}.
\vskip 10pt
\textbf{Acknowledgements:} The work is supported by European Research Council via Synergy Grant 854843 - FASTCORR.
We would like to thank Giuseppe Carleo, Jannes Nys, and Evgeny Stepanov for useful discussions.
We thank SURF~\cite{SURF} for the support in using the National Supercomputer Snellius.
\vskip 10pt
\textbf{Author contributions:} Tom Westerhout suggested the concept of auxiliary sign Ising models, implemented the codes, and ran the simulations.
Andrey Bagrov participated in all project discussions, framed the project and wrote the manuscript.
Mikhail Katsnelson provided general supervision of the research.
All authors helped fine tune the manuscript.
\vskip 10pt
\textbf{Competing interests:} The authors declare no competing interests.

\label{subsec:small}

\end{document}

\renewcommand{\figurename}{Supplementary Figure}

\begin{center}{\Large \textbf{%
    Supplementary Information to the paper ``Many-body quantum sign structures as non-glassy Ising models''
}}\end{center}

\begin{center}
Tom Westerhout,
Mikhail I. Katsnelson,
Andrey A. Bagrov
\end{center}

% \begin{center}
% \today
% \end{center}

\section*{Supplementary Note 1: Reconstructing signs in unfrustrated models}
In the main text, we have stated that, for ground states of unfrustrated quantum systems, the auxiliary classical Ising model is not frustrated either, and that allows one to unambiguously reconstruct the exact ground state sign structure without performing combinatorial optimization.
In what follows, we provide a proof of this statement for $U(1)$-symmetric bi-local spin-$1/2$ Hamiltonians.

Let $\hat{H}$ be such a Hamiltonian. The most general form of a two-site interaction term is then
\begin{gather}
    \hat{H}_{i,j} = %J_{i,j} \left( \sigma^x_i \otimes \sigma^x_j + \sigma^y_i \otimes \sigma^y_j + a^1_{i,j}\cdot \sigma^z_i \otimes \sigma^z_j + a^2_{i,j}\cdot I_i \otimes \sigma^z_j + a^3_{i,j} \cdot \sigma^z_i \otimes I_j+a^4_{i,j} \cdot I_i \otimes I_j \right) = \nonumber  \\
    \left(\begin{array}{cccc}
         a_{i,j} & 0 & 0 & 0 \\
         0 & b_{i,j} & 2J_{i,j} & 0 \\
         0 & 2J_{i,j} & c_{i,j} & 0 \\
         0 & 0 & 0 & d_{i,j}
    \end{array} \right),
\end{gather}
where $\{a,b,c,d\}_{i,j}$ are generic constants that do not enter the couplings of the sign Ising model (and will thus be ignored in the rest of the section), and the only off-diagonal matrix elements $2J_{i,j}$ come from the $J_{i,j}(\sigma^x_i \otimes \sigma^x_j + \sigma^y_i \otimes \sigma^y_j)$ interaction terms that connect Hilbert space basis vectors with flipped spins:
\begin{equation}
    |\dots \uparrow_i \dots \downarrow_j \dots \rangle \longleftrightarrow |\dots \downarrow_i \dots \uparrow_j \dots \rangle.
\end{equation}
If, in the ground state expansion, a basis vector $|a\rangle= |v\rangle \otimes |\uparrow_i \downarrow_j\rangle$ has a sign $\mathcal{S}$ (with $|v\rangle$ defined over all sites except $i$ and $j$), then $|b\rangle=|v\rangle \otimes |\downarrow_i \uparrow_j\rangle$ should have the sign equal to $-\mbox{sign}(\mathcal{J}_{a,b})\cdot \mathcal{S}$, where ${\cal J}_{a,b}=|\psi_a||\psi_b|\langle a| \hat{H}|b\rangle$.
Then, for any closed loop on the Ising model graph, $|s_1\rangle \leftrightarrow |s_2\rangle \leftrightarrow \dots \leftrightarrow |s_M\rangle \leftrightarrow |s_1\rangle$, we can define its parity as 
\begin{equation*}
    P_{\mbox{Ising}} = (-1)^M\prod\limits_{n=1}^{M}
                        \mbox{sign}(\mathcal{J}_{s_{n},s_{n+1}})\,,
                        \;\;\;\mbox{with}\; s_{M+1}=s_1 \,.
\end{equation*}
We say that the Ising model is not frustrated if all non-self-intersecting closed paths have parity $P=1$ (which we call \textit{even}). If this is the case, the sign of a single basis vector unambiguously defines the signs of all of its neighbors, and the full sign structure becomes apparent.

\begin{figure*}[h]
    \centering
    \includegraphics[width=0.6\textwidth]{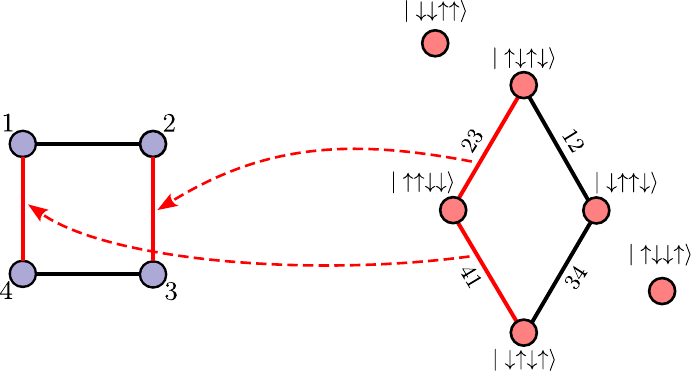}
    
    \caption{\label{fig:loops} {\bf Mapping of a path on the auxiliary classical Ising model onto the physical quantum spin lattice.}
    On the left, we have a 4-site physical lattice, and on the right --- the Hilbert space with a path on the auxiliary Ising model.
    Every edge on the right corresponds to a physical bond on the left connecting two sites.
    A continuous path on the right can result in a discontinuous one on the left as illustrated in red.
    Nevertheless, a closed loop on the Ising graph corresponds to a closed loop of the same length on the physical lattice.}
\end{figure*}

Note, that each closed path on the auxiliary Ising model graph corresponds to a closed path of the same length on the physical lattice (see Sup.~Fig.~\ref{fig:loops}). Each edge in the Ising model corresponds to an exchange of spins at some positions $i$ and $j$, and it maps to the $i,j$-bond of the physical lattice (multiple Ising links can be mapped to the same physical link). By moving along a closed non-self-intersecting path, one flips every spin exactly two times, which implies that the corresponding bonds on the physical lattice form a closed path of the same length.

Parity of the loop on the physical lattice,
\begin{equation*}
    P_{\mbox{quantum}} = (-1)^M \prod\limits_{n=1}^{M} \mbox{sign}(J_{i_n,i_{n+1}})\,, \;\;\mbox{with}\; i_{M+1}=i_1 \,,
\end{equation*}
is then exactly the same as $P_{\mbox{Ising}}$, because $\mbox{sign} (J_{i,j})= \mbox{sign}({\cal J}_{a,b})$ for $|a\rangle$ and $|b\rangle$ that differ only by the exchange of spins at positions $i$ and $j$.

Hence, if any closed path on the physical lattice of the quantum model has even parity, the resulting auxiliary sign Ising model lacks frustrations, and the pattern of signs can be straightforwardly reconstructed.
For instance, for the Heisenberg antiferromagnet on a bipartite lattice, the auxiliary classical sign Ising model is also bipartite and antiferromagnetic, and the alternating pattern of signs is trivially reconstructed.
However, this statement remains correct for any bi-local $U(1)$-invariant spin-$1/2$ quantum magnet where every closed loop on its lattice has even parity $P=1$.
For example, the Heisenberg magnet on the triangular lattice with ferromagnetic exchange along one axis and antiferromagnetic exchange along the other two axes.

\section*{Supplementary Note 2: Physical systems for numerical simulations}

Numerical simulations were performed on small- to medium-sized clusters that were still amenable to exact diagonalization. In ``Small quantum systems'' subsection of the main text we considered the following models:

\begin{figure*}[h]
    \centering
    \includegraphics[width=\textwidth]{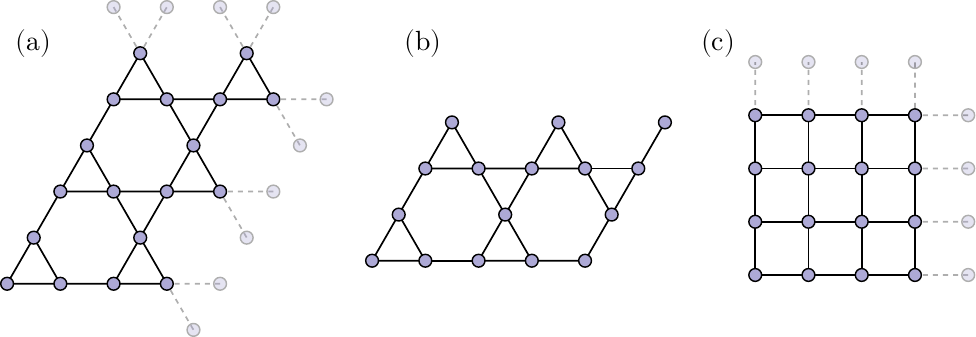}
    
    \caption{\label{fig:small_physical_systems}{\bf Some of the studied Heisenberg models.} (a) 18-site Kagome lattice, (b) 16-site Kagome lattice, and (c) 16-site square lattice. Periodic boundaries in 2 directions are indicated with dashed lines.}
\end{figure*}

\begin{itemize}
    \item Heisenberg model on 18-site Kagome lattice with periodic boundary conditions (see Sup.~Fig.~\ref{fig:small_physical_systems}(a));
    \item Heisenberg model on 16-site Kagome lattice with open boundary conditions (see Sup.~Fig.~\ref{fig:small_physical_systems}(b));
    \item $J_1$-$J_2$ model on 16-site square lattice with periodic boundary conditions (see Sup.~Fig.~\ref{fig:small_physical_systems}(c)), $J_2/J_1$ ratio was set to $0.55$;
    \item 16-site all-to-all connected model with random Heisenberg interactions $J_{ij} \sim \mathrm{Normal}(\mu=0, \sigma=1)$.
\end{itemize}

For all of these models we restricted the Hilbert space to the sector of zero magnetization.

In Fig.~6 of the main text we considered a few larger models, namely:

\begin{itemize}
    \item Heisenberg model on 32-site Pyrochlore cluster~\cite{astrakhantsev2021broken};
    \item Heisenberg model on 36-site Kagome lattice~\cite{leung1993numerical};
    \item 32-site all-to-all connected model with random Heisenberg interactions\\ ${J_{ij} \sim \mathrm{Normal}(\mu=0, \sigma=1)}$.
\end{itemize}

For both 16- and 32-site random Heisenberg models we chose the interactions $J_{ij}$ to be distributed according to Normal (Gaussian) distribution with mean $\mu=0$ and standard deviation $\sigma=1$. However, changing $\sigma$ would only result in energy rescaling and would not affect our results.

\section*{Supplementary Note 3: Details of running the Simulated Annealing algorithm on small systems}

In this section, we provide some more details about running SA algorithm~\cite{kirkpatrick1983optimization} that was used to produce Fig.~2 in the main text.
Specifically, we have implemented the algorithm~\cite{ising-glass-annealer} using tricks from~\cite{isakov2015optimised} to improve its performance.

We determine the initial and final annealing temperatures automatically according to
\begin{equation*}
\begin{aligned}
    \beta_\mathrm{min} &= \log 2 / \Delta E_\mathrm{max} \,, \\
    \beta_\mathrm{max} &= \log 100 / \Delta E_\mathrm{min} \,.
\end{aligned}
\end{equation*}
Here, $\Delta E_\mathrm{max}$ and $\Delta E_\mathrm{min}$ are maximal and minimal energy changes which can result from a single spin flip.
We refer the reader to our implementation~\cite{ising-glass-annealer} of the algorithm that estimates $\Delta E_\mathrm{max}$ and $\Delta E_\mathrm{min}$. 
In all simulations, we use exponential annealing schedule because it produced better results than linear.
The exponential schedule is defined by the following equation:
\begin{equation*}
    \beta(t) = \beta_\mathrm{min} \left(\frac{\beta_\mathrm{max}}{\beta_\mathrm{min}}\right) ^ {t / (T - 1)} \,,\,\, t=0\dots T-1
\end{equation*}
where $T$ is the total number of sweeps.

For every model, we perform 10 runs with difference random number generator seeds (to roughly estimate the statistical errors).
Within each run, for a given number of sweeps, we perform SA 1024 times to estimate the probability of obtaining $99.5\%$ accuracy.

\vskip 10pt
\section*{Supplementary Note 4: Robustness of the algorithm to noise in amplitudes}
\begin{figure}[h!]
    \centering
    \includegraphics[width=0.7\columnwidth]{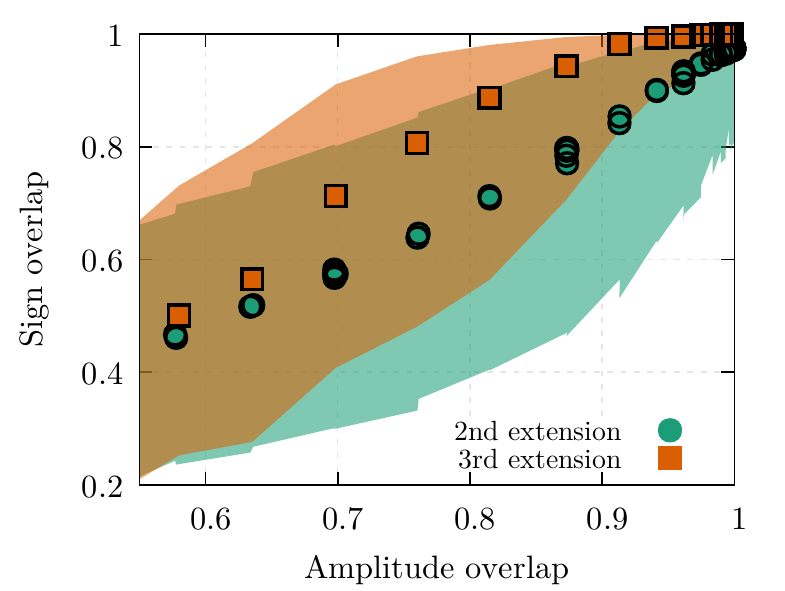}
    \caption{{\bf Effect of the noise in the amplitudes on the quality of the reconstructed sign structure for random connected clusters $\mathcal{C}$.} The experiments were performed on the 36-site Kagome lattice. We show the sign overlap $O_{\cal C}^{\cal S}$ as a function of the amplitude overlap $O_{\cal C}^{\cal A}$. Green circles depict median values of the sign overlap computed using ${\cal C}^{(2)}$ extensions of the randomly sampled connected clusters, and orange squares --- the median values computed using ${\cal C}^{(3)}$ extensions. Shaded regions show the interquartile range of $O_{\cal C}^{\cal S}$. For each point on the plot, $6000$ clusters of sizes $50$--$1000$ have been sampled to collect statistics.}
    \label{fig:kagome_36_noise}
\end{figure}

To demonstrate the robustness of the algorithm with respect to noise in the amplitudes for larger systems, where greedy optimization cannot be performed on the full Hilbert space basis, we use the ground state of the 36-site Kagome antiferromagnet and corrupt its amplitudes in the same way as outlined in the main text.
Then, we randomly sample clusters $\cal C$, construct their extensions ${\cal C}^{(2)}$ and ${\cal C}^{(3)}$, and use the greedy algorithm to retrieve sign structures on $\cal C$. From Sup.~Fig.~\ref{fig:kagome_36_noise}, one can see that, for larger systems, stability of the algorithm is less pronounced than for smaller systems. This is likely because of the interplay between errors resulting from the noise in the amplitudes and errors corresponding to uncertainty in the ``external magnetic fields'' $h_i$ at the boundary of the cluster extension. However, if $O_{\cal C}^{\cal A} >0.9$, optimization on ${\cal C}^{(2)}$ leads to good, and on ${\cal C}^{(3)}$ --- to nearly perfect results. It should be noted that $O_{\cal C}^{\cal A} = 0.9$ corresponds to the noise level $\epsilon \approx 0.8$, and the maximal amplitude of the ground state, $\max_i |\psi^{exact}_i|= 0.017$, acquires corrupted values in the range $(0.007,0.038)$.
\vskip 10pt

\section*{Supplementary Note 5: Sign optimization within variational Monte Carlo loop}
Here, we outline how the approach to reconstructing sign structures via combinatorial optimization of the auxiliary classical Ising model can be used in conjunction with variational Monte Carlo (vMC) algorithms to solve larger-scale problems. Consider a variational ansatz:
\begin{equation}
    \psi_W = \sum\limits_{i} \psi({\bf W}, i) | i \rangle,
\end{equation}
where $\bf W$ are variational parameters of the ansazt (e.g., neural network weights in case $\psi_W$ is an NQS), $| i \rangle = | \dots \uparrow \dots \downarrow \dots \rangle$ are basis vectors of the Hilbert space, and the sum runs over the whole basis.

Traditionally, within the vMC framework, the optimal approximation to the quantum Hamiltonian $\hat{H}$ ground state is obtained through step by step minimization of the energy of the variational ansatz:
\begin{gather}
E_n = \frac{\langle \psi_{W_n} | \hat{H} | \psi_{W_n} \rangle}{\langle \psi_{W_n} | \psi_{W_n} \rangle}, \\
{\bf W}_{n+1} = {\bf W}_{n} - \gamma \frac{\partial E_n}{\partial \mathbf{W}},
\end{gather}
where $n$ is the current optimization step, and $\gamma$ is the learning rate.

The Hamiltonian expectation value is evaluated by means of Monte Carlo sampling as 
\begin{equation}
     E_W = \sum\limits_{i} \frac{\langle i|\hat{H}|\psi_W \rangle}{ \langle i|\psi_W \rangle} \left|\frac{\langle i|\psi_W \rangle}{\langle \psi_W| \psi_W\rangle}\right|^2 \simeq \frac{1}{K} \sum\limits_{i=i_1}^{i_K} \frac{\langle i|\hat{H}|\psi_W \rangle}{ \langle i|\psi_W \rangle} \equiv \frac{1}{K} \sum\limits_{i=i_1}^{i_K} E_{loc}(i),
\end{equation}
where $K$ is the number of basis vectors sampled from probability distribution $p_W(i) = \left|\frac{\langle i|\psi_W \rangle}{\langle \psi_W| \psi_W\rangle}\right|^2$, which typically scales linearly with the size of the system, $K \sim 2000 \cdot N$.

More formally, the optimization algorithm can be outlined as follows:

\normalem
\begin{algorithm}[H]\label{alg:plain-mc}
    \DontPrintSemicolon
    ${\bf W}_0 \leftarrow \text{random weights}$ \;
    $n \leftarrow 0$ \;
    
    \While{$\mathrm{not}\;\mathrm{converged}$}{
        $\{|i_1\rangle, \dots |i_K\rangle \} \leftarrow p_{W_n}(i)$ \;
        $E_n \leftarrow \displaystyle{\frac{1}{K} \sum\limits_{i=i_1}^{i_K} \frac{\langle \psi_{W_n}|\hat{H}|i \rangle}{ \langle \psi_{W_n}| i\rangle}}$ \;
        ${\bf W}_{n+1} \leftarrow {\bf W}_{n} - \displaystyle{\gamma \frac{\partial E_n}{\partial \mathbf{W}_n}}$ \;
        $n \leftarrow n+1$ \;
    }
\end{algorithm}
\ULforem

Here, the variational ansatz $\psi_W$ encodes the wave function coefficients --- both their amplitudes and phases/signs: $\langle \psi_W| i\rangle~=~\psi({\bf W}, i)$. To implement our idea of combinatorial optimization of sign structure, we shall instead use variational representation only for the amplitudes, which we denote as $A({\bf W}, i)$, while storing signs ${\cal S}_i$ of relevant basis vectors explicitly: $\langle \psi_W| i\rangle = A({\bf W}, i){\cal S}_i$.

The Monte Carlo estimate of the energy of the combined representation becomes
\begin{equation}
    E_W \simeq \frac{1}{K} \sum\limits_{i=i_1}^{i_K} E_{loc}(i) = \frac{1}{K} \sum\limits_{i=i_1}^{i_K} \frac{\sum_j A({\bf W}, j) {\cal S}_j H_{ij}}{A({\bf W}, i) {\cal S}_i}. 
\end{equation}
At every optimization step, one needs the signs of basis vectors $| i \rangle \in \left\{| i_1 \rangle,\dots |i_K \rangle \right\}$, as well as of all the vectors $|j\rangle$ for which $\langle j | \hat{H} | i \rangle = H_{ij} \neq 0$ for some $|i \rangle$. Let us denote a vector $| i \rangle$ together with its Hamiltonian neighbours $\left\{ | j \rangle: H_{ij}\neq0\right\}$ as $M_i$, and the complete set of vectors for which the signs are to be reconstructed as 
\begin{equation}
M = \bigcup\limits_{i=i_1}^{i_K} M_i
\end{equation}

The optimization algorithm then acquires the form:

\normalem
\begin{algorithm}[H]\label{alg:plain-mc}
    \DontPrintSemicolon
    ${\bf W}_0 \leftarrow \text{random weights}$ \;
    $n \leftarrow 0$ \;
    
    \While{$\mathrm{not}\;\mathrm{converged}$}{
        $\{|i_1\rangle, \dots |i_K\rangle \} \leftarrow p_{W_n}(i)$ \;
        $M \longleftarrow {\cal C}^{(1)} (\{|i_1\rangle, \dots |i_K\rangle \})$ \;
        $\{{\cal S}_{i}: i \in M\} \leftarrow \mbox{Greedy}({\cal C}^{(2)}(M))$ \;
        $E_{W_n} \leftarrow \displaystyle{\frac{1}{K} \sum\limits_{i=i_1}^{i_K} \frac{\sum_j A({\bf W}_n, j) {\cal S}_j H_{ij}}{A({\bf W}_n, i) {\cal S}_i}}$ \;
        % Compute $\partial_W E_{W_n}$ with fixed $\{{\cal S}_{i}: i \in M\}$ \;
        % ${\bf W}_{n+1} \leftarrow {\bf W}_{n} - \gamma \partial_W E_{W_n}$ \;
        \tcp{In the following, $\{{\cal S}_{i}: i \in M\}$ are kept fixed during differentiation}
        ${\bf W}_{n+1} \leftarrow {\bf W}_{n} - \displaystyle{\gamma \frac{\partial}{\partial \mathbf{W}_n} E_n}$ \;
        $n \leftarrow n+1$ \;
    }
\end{algorithm}
\ULforem

Line 6 of this loop requires elaboration. In the main text of the paper, we have shown how greedy optimization can be used to find signs of connected (and not too large) clusters in the Hilbert space basis. Here, set $M$ is different from those clusters in two ways. First, with high probability this set itself, as well as its extension ${\cal C}^{(2)}(M)$ (or a higher order one), have many disconnected components. Secondly, it can be pretty large. For example, if the underlying quantum Heisenberg Hamiltonian is defined on 
a regular lattice with coordination number $z$, every basis vector $|i\rangle$ has up to $zN$ neighbours, and the size of $M$ is
\begin{equation}
    \#M < zNK \sim 8 \cdot 10^7
\end{equation}
for $K=2000\cdot N$, $z=4$, $N=100$. Its second extension would have
\begin{equation}
    \#{\cal C}^{(2)}(M) < \#M \cdot z^2 N^2 < 1.28 \cdot 10^{13}.
\end{equation}
In practice, this upper bound is never reached because many vectors share common neighbours (and the actual size of $M$ is about two orders of magnitude smaller), but anyway the resulting set is way too large to run combinatorial optimization on it.

To prove applicability of the combinatorial approach to retrieving the signs, we shall now show two things: 1) optimization of signs can be performed on each disconnected component of ${\cal C}^{(2)}(M)$ independently; 2) with probability approaching $1$, {\it all} extensions ${\cal C}^{(2)}(M_i)$ are disconnected from each other, and the optimization algorithm can be executed for each of $K$ small independent Ising subsystems.

The first point can be proven by simply noticing that, to correctly compute $E_{loc}(i)$, one needs to know only the relative signs of $|i \rangle$ and $| j \rangle$, and a flip of both ${\cal S}_i$ and $\{{\cal S}_j\}$ does not change the value of $E_{loc}(i)$:
\begin{equation}
    E_{loc}(i) = \frac{\sum_j A({\bf W}, j) {\cal S}_j H_{ij}}{A({\bf W}, i) {\cal S}_i} = \frac{\sum_j A({\bf W}, j) ({-\cal S}_j) H_{ij}}{A({\bf W}, i) ({-\cal S}_i)}.
\end{equation}
This implies that, if two sampled basis vectors $|i_1 \rangle$ and $|i_2 \rangle$ have non-intersecting extensions of their neighbourhoods $M_{i_1}$ and $M_{i_2}$: ${\cal C}^{(2)}(M_{i_1}) \cap {\cal C}^{(2)}(M_{i_2})=\emptyset$, their signs structures can be optimized independently.
Though as a result of independent optimization, they may acquire incorrect relative sign, each of the two sign structures is nevertheless a good approximation to the true one (which is also defined up to an overall flip) and gives the correct result for $E_{loc}(i_1)$ or $E_{loc}(i_2)$ correspondingly.

However, if two or more extensions intersect, their joint sign structure must be optimized as a whole. Hence, the combinatorial optimization of signs would be possible only if large clusters unifying many extensions ${\cal C}^{(2)}(M_i)$ do not generally form.

This can be proven by constructing the lower bound on the probability $P$ that the following statement holds:
\begin{equation}
\forall \,\, i\neq j \in \left\{ i_1,\dots i_K\right\}: \;\; {\cal C}^{(2)}(M_i) \cap {\cal C}^{(2)}(M_j) = \emptyset.
\end{equation}
Let us sample basis vectors $|i_1\rangle, |i_2\rangle, |i_3\rangle, \dots$ from the uniform distribution. Once the first vector $|i_1\rangle$ is sampled (then $M_{i_1}={\cal C}^{(1)}(|i_1\rangle)$), the second sampled vector must not belong to ${\cal C}^{(6)}(|i_1\rangle)$ to ensure that ${\cal C}^{(2)}(M_{i_1}) \cap {\cal C}^{(2)}(M_{i_2}) = \emptyset$. The size of ${\cal C}^{(6)}(|i_1\rangle)$ is bounded from above by $z^6 N^6$ (much smaller in practice), so the probability of this event is 
\begin{equation}
    p ({\cal C}^{(2)}(M_{i_1}) \cap {\cal C}^{(2)}(M_{i_2}) = \emptyset) \geq 1 - \frac{z^6 N^6}{{\cal C}^N_{N/2}}.
\end{equation}
When the third vector is sampled, it must not belong to ${\cal C}^{(6)}(|i_1\rangle) \bigcup {\cal C}^{(6)}(|i_2\rangle)$, which has a maximal size of $2z^6N^6$, etc. The lower bound on the probability that all sampled vectors produce disconnected clusters in the Hilbert space basis, and their sign structures can be optimized independently, is then
\begin{equation}
    P > \prod\limits_{n=1}^{K} (1-\frac{nz^6N^6}{{\cal C}^N_{N/2}}) > (1-\frac{K z^6N^6}{{\cal C}^N_{N/2}})^K.
\end{equation}
E.g., for $z=4$, $N=100$, $K=2\cdot 10^5$, it gives
\begin{equation}
    P > (1-8.12\cdot 10^{-9})^{200000} \simeq 0.9983,
\end{equation}
and it should be kept in mind that formation of small connected components unifying several $M_i$ sets does not cause problems for the greedy algorithm.

From here, the ${\cal O} (N^4 \log N)$ complexity scaling can deduced. Sizes of individual connected ${\cal C}^{(2)}(M_i)$ clusters scale as ${\cal O} (N^3)$ (because $\# M_i =zN$), with the number of clusters $K$ being linear in $N$.
Total runtime of the greedy algorithm then amounts to $K {\cal O} (N^3 \log N^3) \sim {\cal O} (N^4 \log N)$.

In practice, the sampling occurs from the probability distribution set by the wavefunction amplitudes, $p(i) \sim |A({\bf W},i)|^2$, which is not uniform, but if this leads to formation of unwanted large connected clusters, those can be broken up by flattening the probability distribution through importance sampling.

\section*{Supplementary Note 6: Reconstructing sign structure on random clusters with Simulated  Annealing}

\begin{figure*}[h!]
    \centering
    \includegraphics[width=0.95\textwidth]{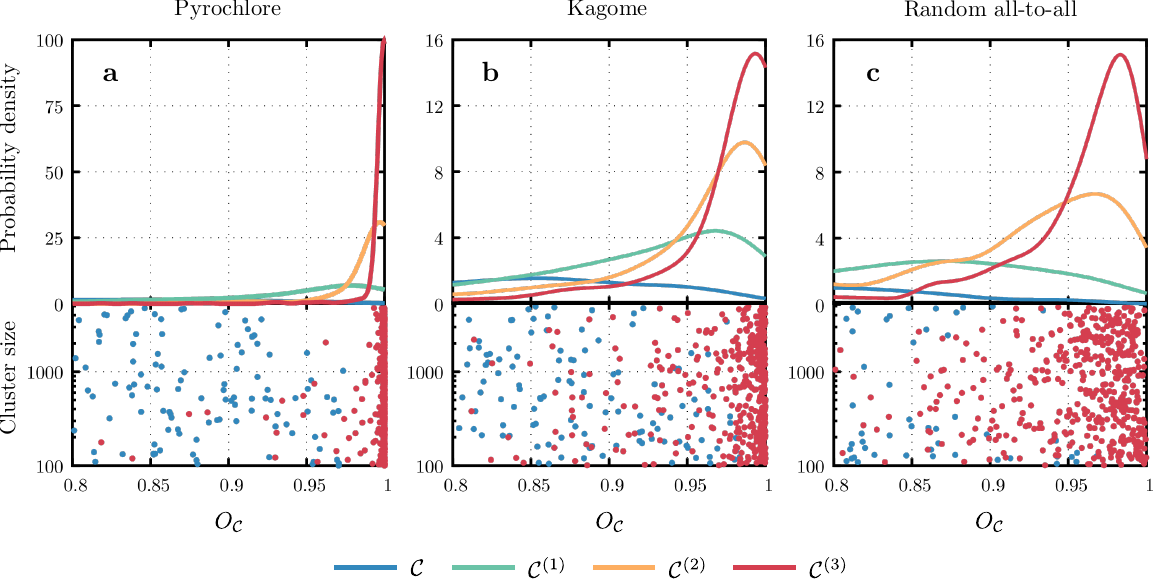}
    \caption{{\bf Quality of the sign structures reconstructed using SA on random small connected clusters.} The results are shown for \textbf{a} the 32-site pyrochlore lattice, \textbf{b} 36-site Kagome lattice, and \textbf{c} 32-site all-to-all connected model with random couplings. 
    The top panel shows the distribution (or the probability density function) of clusters as a function of the overlap $O_\mathcal{C}$.
    We run SA on the original cluster $\mathcal{C}$ as well as on its Hamiltonian extensions ${\cal C}^{(1)}$, ${\cal C}^{(2)}$, and ${\cal C}^{(3)}$.
    The bottom panel shows the raw data from which the probability densities were estimated.
    Only $\mathcal{C}$ and $\mathcal{C}^{(3)}$ are shown to reduce the visual noise.
    There is no correlation between the overlap $O_\mathcal{C}$ and the cluster size.}
    \label{fig:overlap_and_scatter} 
\end{figure*}

In the main text, we relied on the greedy algorithm to reconstruct signs on random small connected clusters sampled from the Hilbert space basis of several middle-size quantum models -- the Heisenberg antiferromagnet on 32-site pyrochlore lattice, 36-site Kagome lattice, and 32-site fully connected graph with random couplings. In Sup.~Fig. \ref{fig:overlap_and_scatter}, similar results obtained with SA algorithm are shown.

We considered $500$ random clusters.
Their sizes were distributed logarithmically between $100$ and $5000$.
For each cluster and each of its Hamiltonian extensions, we ran simulated annealing $64$ times using $5000$ sweeps.
The probability density in the top panel was obtained from the raw data using the Gaussian kernel density estimation algorithm.

\section*{Supplementary Note 7: Extension to not time-reversal-symmetric Hamiltonians}

If the quantum lattice model is not time-reversal symmetric, its Hamiltonian is inherently complex-valued.
In that case, the phase structure cannot be reduced to binary signs, and the suggested mapping onto the classical Ising model is impossible.
Instead, one can reconstruct the phase structure by optimizing the energy of a continuous rotor model, also known as the XY model.
Indeed, the quantum state is
\begin{gather}
    |\psi \rangle = \sum\limits_i |\psi_i| e^{i \phi_i} |i \rangle,
\end{gather}
and the corresponding energy is
\begin{gather}
    E = \langle \psi | \hat{H} | \psi \rangle =
    \sum\limits_{i,j} \langle i | \hat{H} | j \rangle |\psi_i||\psi_j| e^{i(\phi_i - \phi_j)} = \\
    \sum\limits_{i,j} \frac{1}{2}\left( \langle i | \hat{H} | j \rangle e^{i(\phi_i - \phi_j)} + \langle j | \hat{H} | i \rangle e^{i(\phi_j - \phi_i)} \right) \cdot |\psi_i||\psi_j| \nonumber = \\
    \sum\limits_{i,j} \text{Re}\left[ \langle i | \hat{H} | j \rangle e^{i(\phi_i - \phi_j)} \right] \cdot |\psi_i||\psi_j| \nonumber = \\
    \sum\limits_{i,j} \mathcal{J}_{i,j} \cos \left(\Delta_{i,j} + \phi_i - \phi_j \right), \nonumber
\end{gather}
where we have expressed the Hamiltonian matrix elements as $\langle i | H | j\rangle = |H_{i,j}|\cdot e^{i\Delta_{i,j}}$ and have defined $\mathcal{J}_{i,j} = |H_{i,j}|\cdot |\psi_i||\psi_j|$.

\bibliography{92_references_supplement.bib}